\documentstyle[preprint,prd,epsf,floats,aps]{revtex}

\def\quatrefigures#1#2#3#4#5#6{
\begin{figure} 
\centerline{
   \begin{tabular}{c@{\hspace{1cm}}c}
           \mbox{\epsfxsize=7cm \epsfbox{#1} }&
           \mbox{\epsfxsize=7cm \epsfbox{#2} }\\(a)&(b)\\[5mm]
           \mbox{\epsfxsize=7cm \epsfbox{#3} }&
           \mbox{\epsfxsize=7cm \epsfbox{#4} }\\(c)&(d)
   \end{tabular}
}
\caption{#5}
\label{#6}
\end{figure}
}

\def\duesfigures#1#2#3#4{  
\begin{figure} 
\centerline{
   \begin{tabular}{c@{\hspace{1cm}}c}
           \mbox{\epsfxsize=7cm \epsfbox{#1} }&
           \mbox{\epsfxsize=7cm \epsfbox{#2} }\\(a)&(b)
   \end{tabular}
}
\caption{#3}
\label{#4}
\end{figure}
}

\def\unafigura#1#2#3#4{  
\begin{figure} 
\centerline{\mbox{\epsfxsize=#4 \epsfbox{#1} }}
\caption{#2}
\label{#3}
\end{figure}
}

\newcommand{\figplane}{
\unafigura{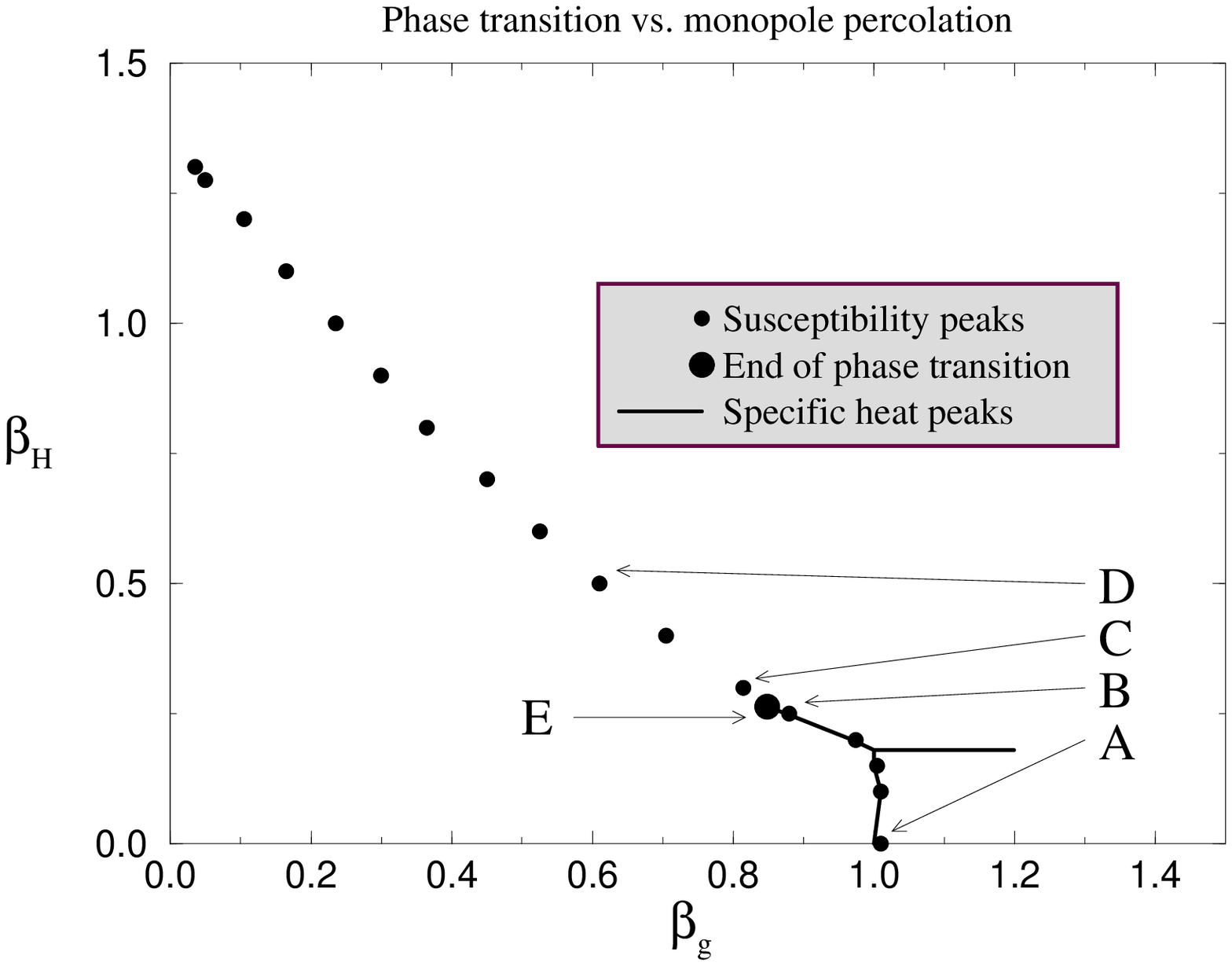}
          {The plane $(\beta_g, \beta_H)$ showing the lines of
          phase transition and the monopole percolation.
          Thermal cycles have been done at fixed $\beta_H$ values:
          A~($\beta_H=0.00$),
          B~($\beta_H=0.25$),
          C~($\beta_H=0.30$) and
          D~($\beta_H=0.30$).
          The magnitudes measured along this lines are shown in the 
          Figs.~\ref{fig:dades}a,~\ref{fig:dades}b,~\ref{fig:dades}c
          and~\ref{fig:dades}d respectively. The point E is the end of the
          phase transition.
          }{fig:plane}{8cm}
}

\newcommand{\figpeaks}{
\unafigura{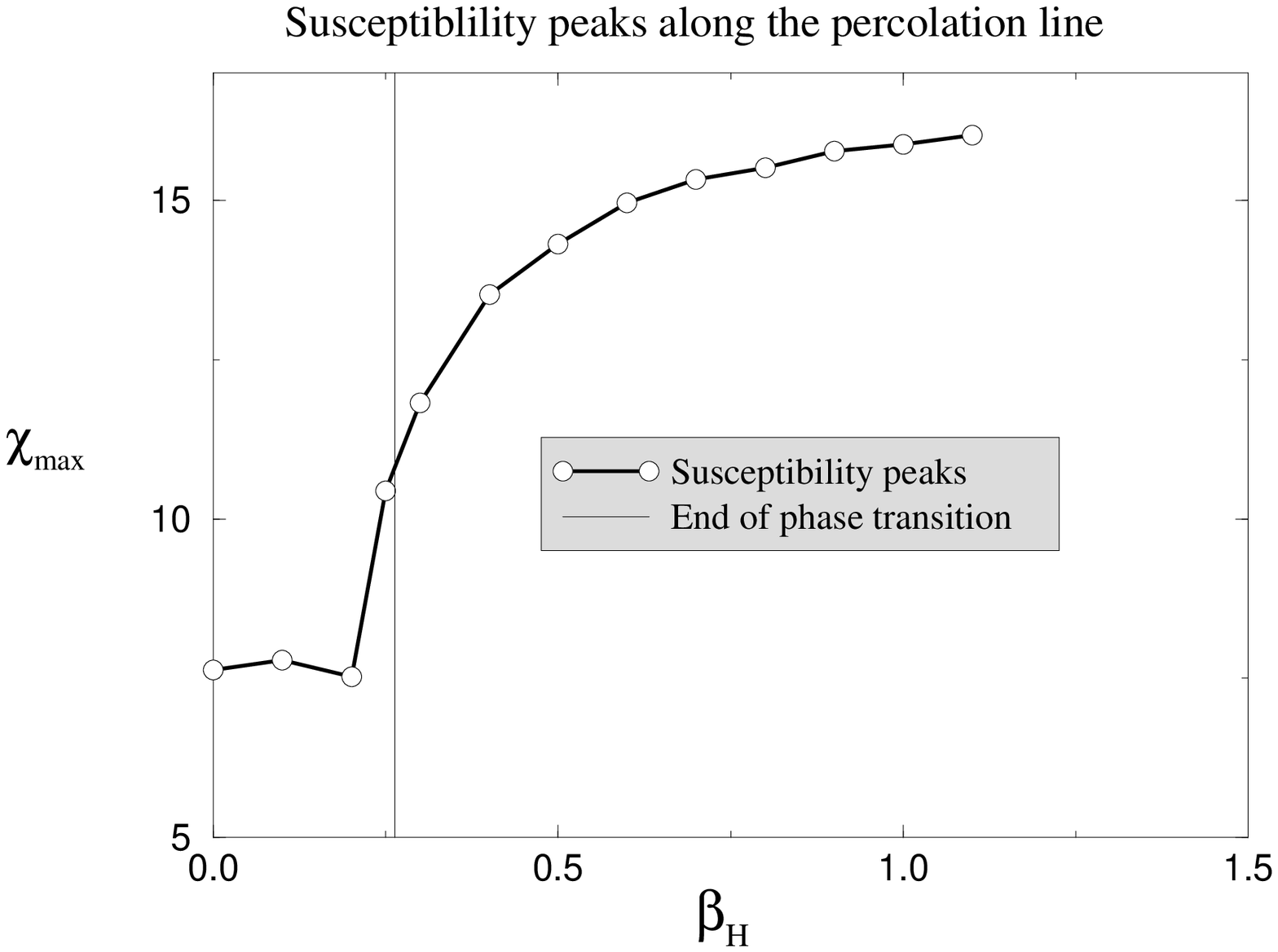}
          {Susceptibility peaks along the percolation line. Notice the 
           change of behavior at the end of the phase transition.}
          {fig:peaks}{6cm}
}

\newcommand{\figdades}{
\quatrefigures{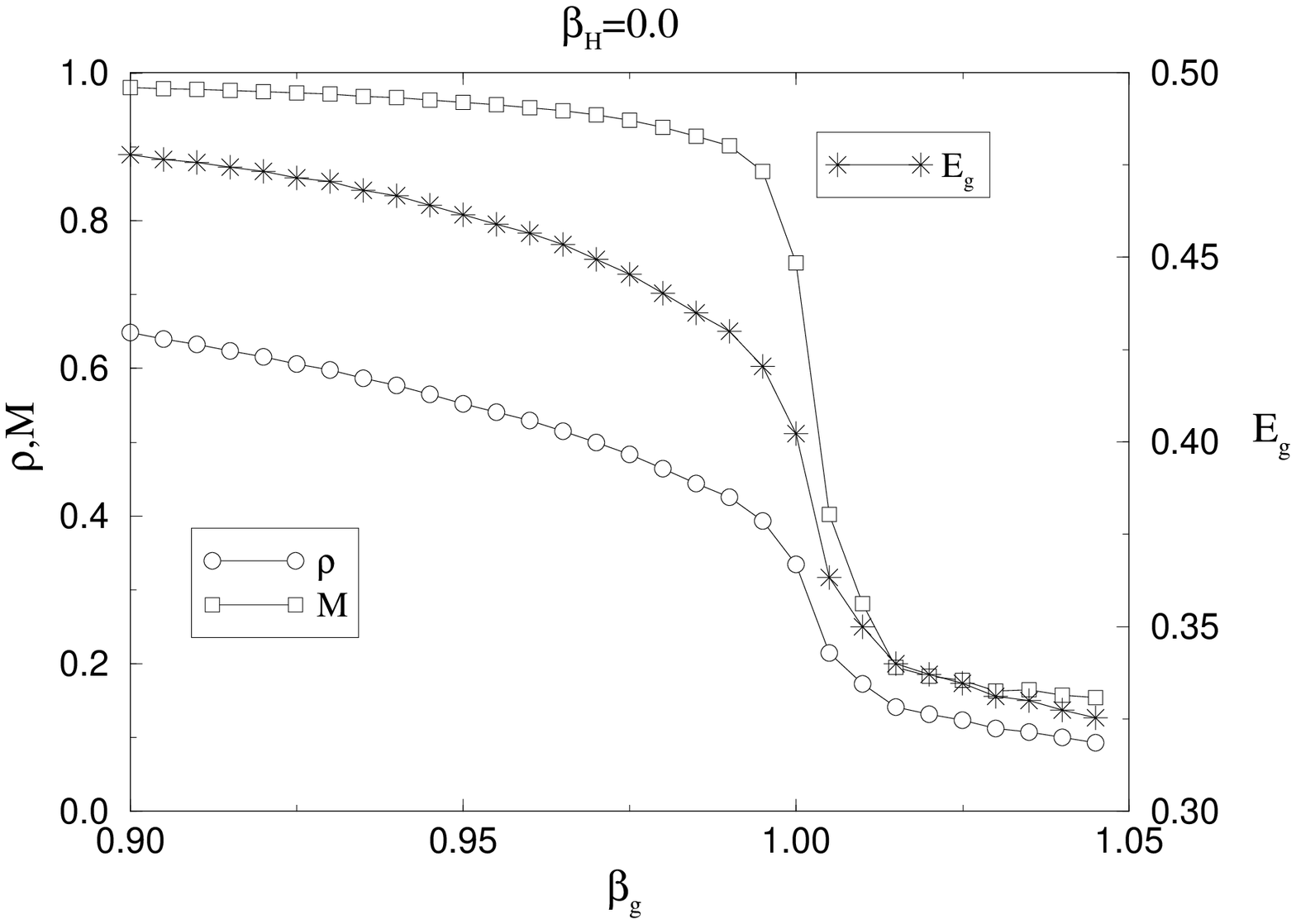}{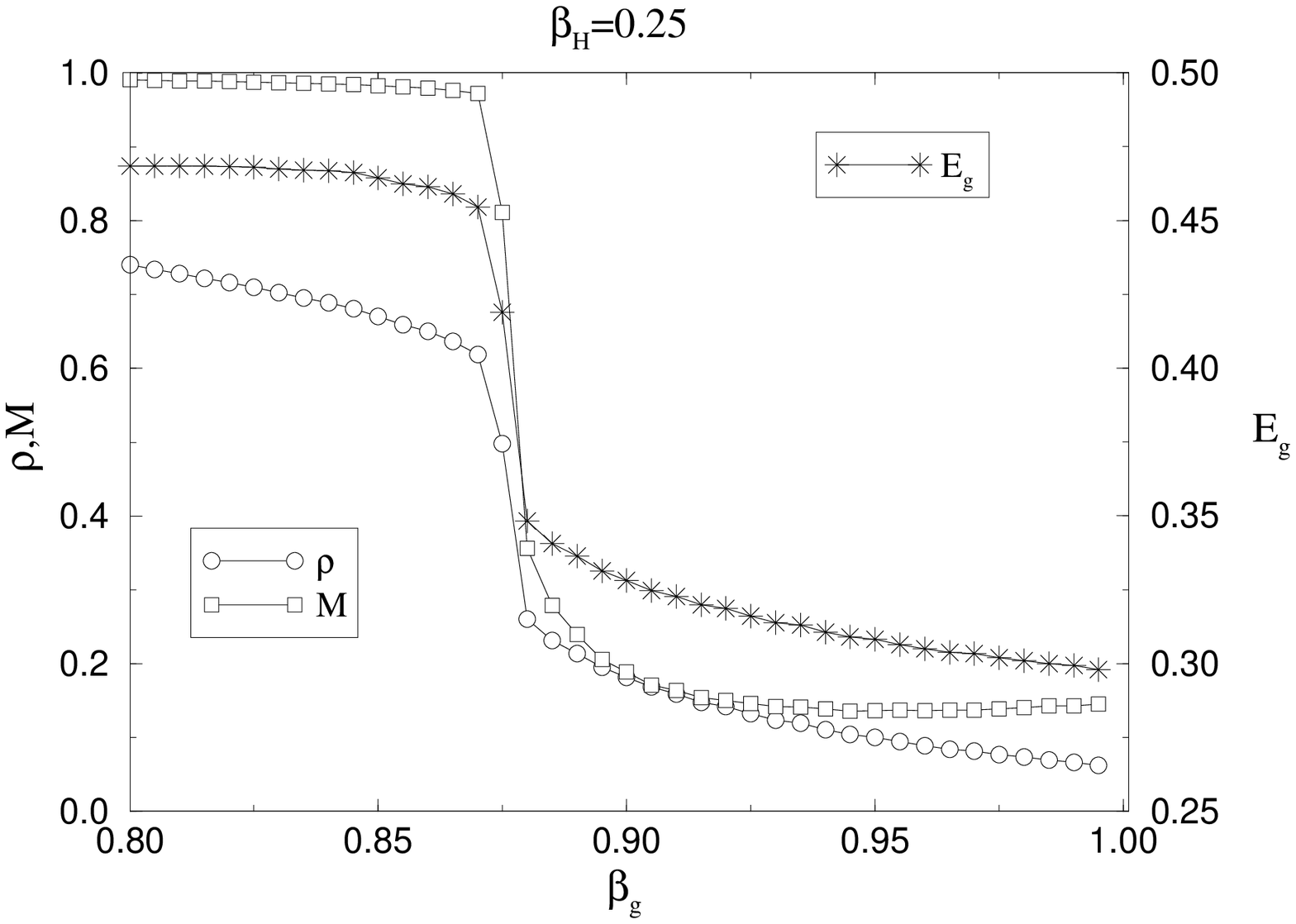}{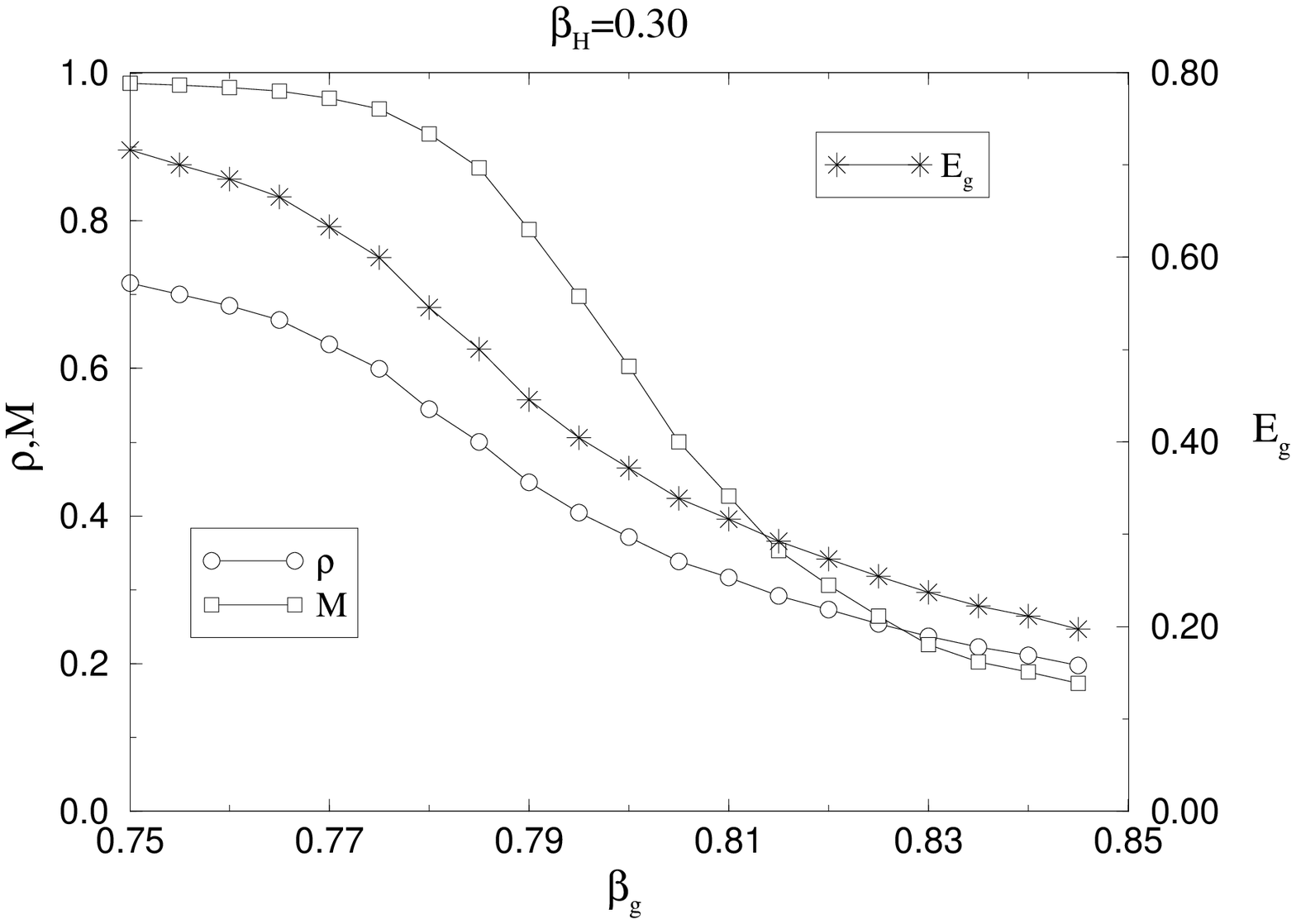}{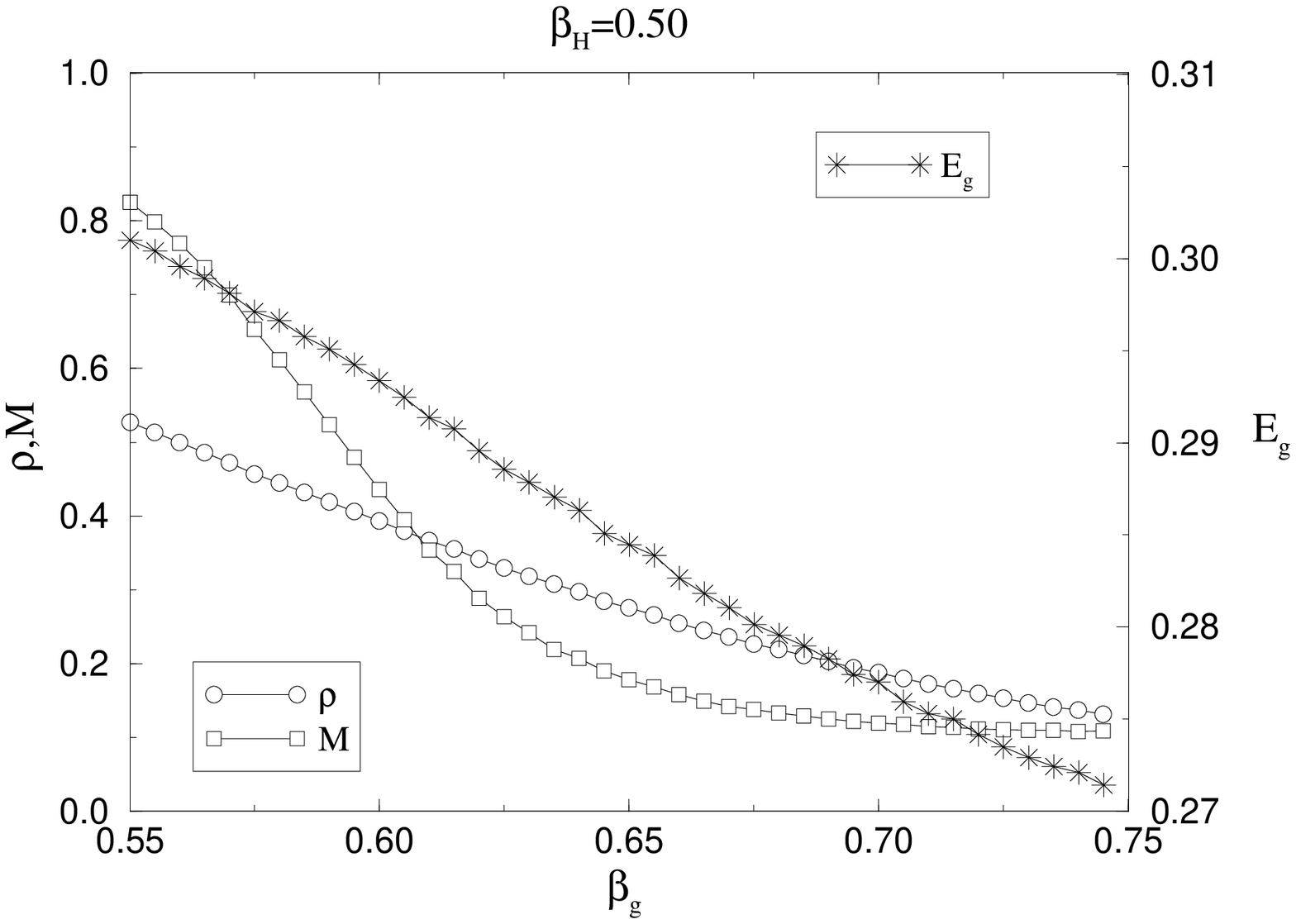}
              {Results for thermal cycles at different 
              $\beta_H$ (points A, B, C and D in the Fig.~\ref{fig:plane}).
              The values of $\rho$ and $M$ can be read 
              in the left $Y$ axes, and the energies in the right $Y$ axes.
              }
              {fig:dades}
}

\newcommand{\fighisto}{
\duesfigures{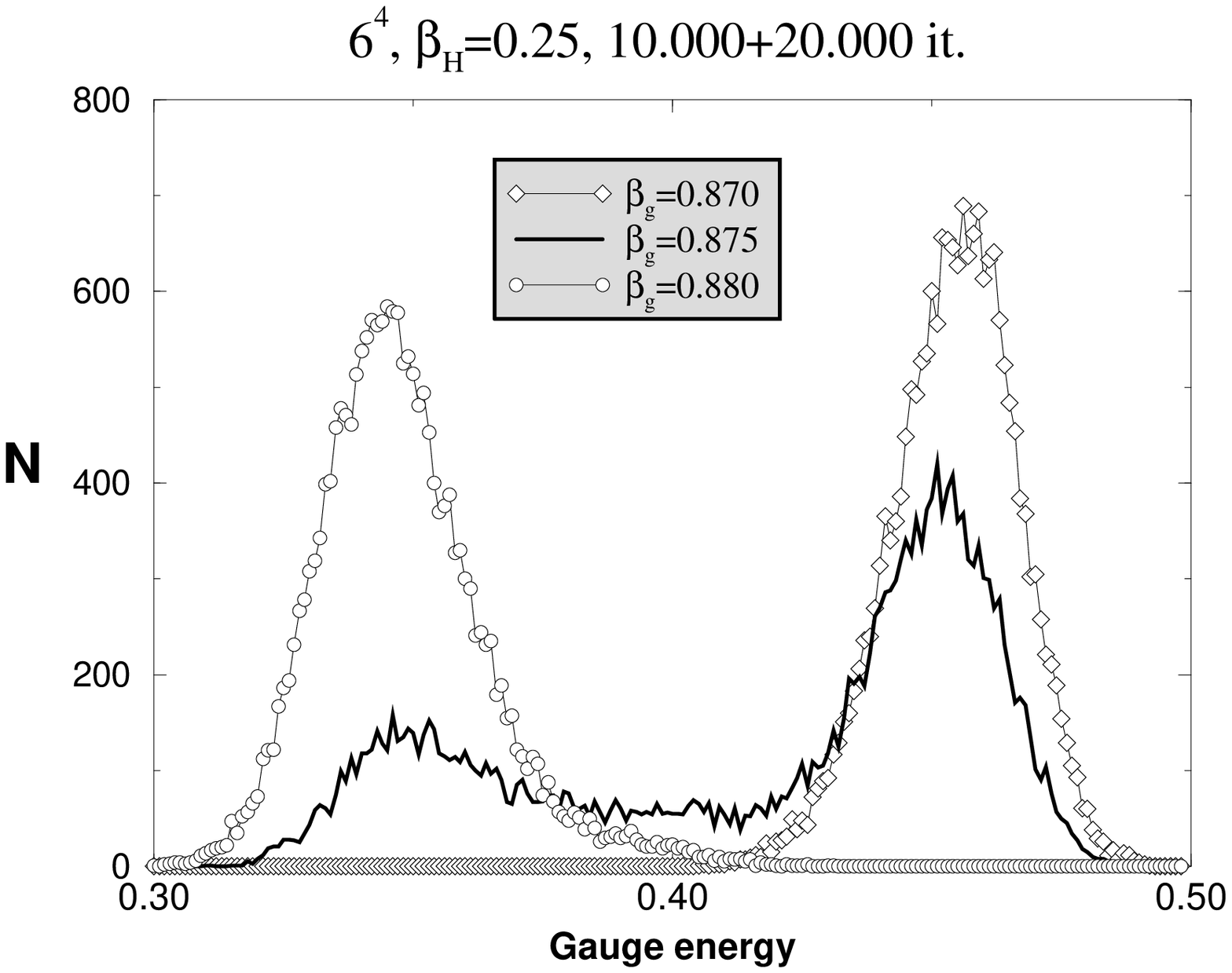}{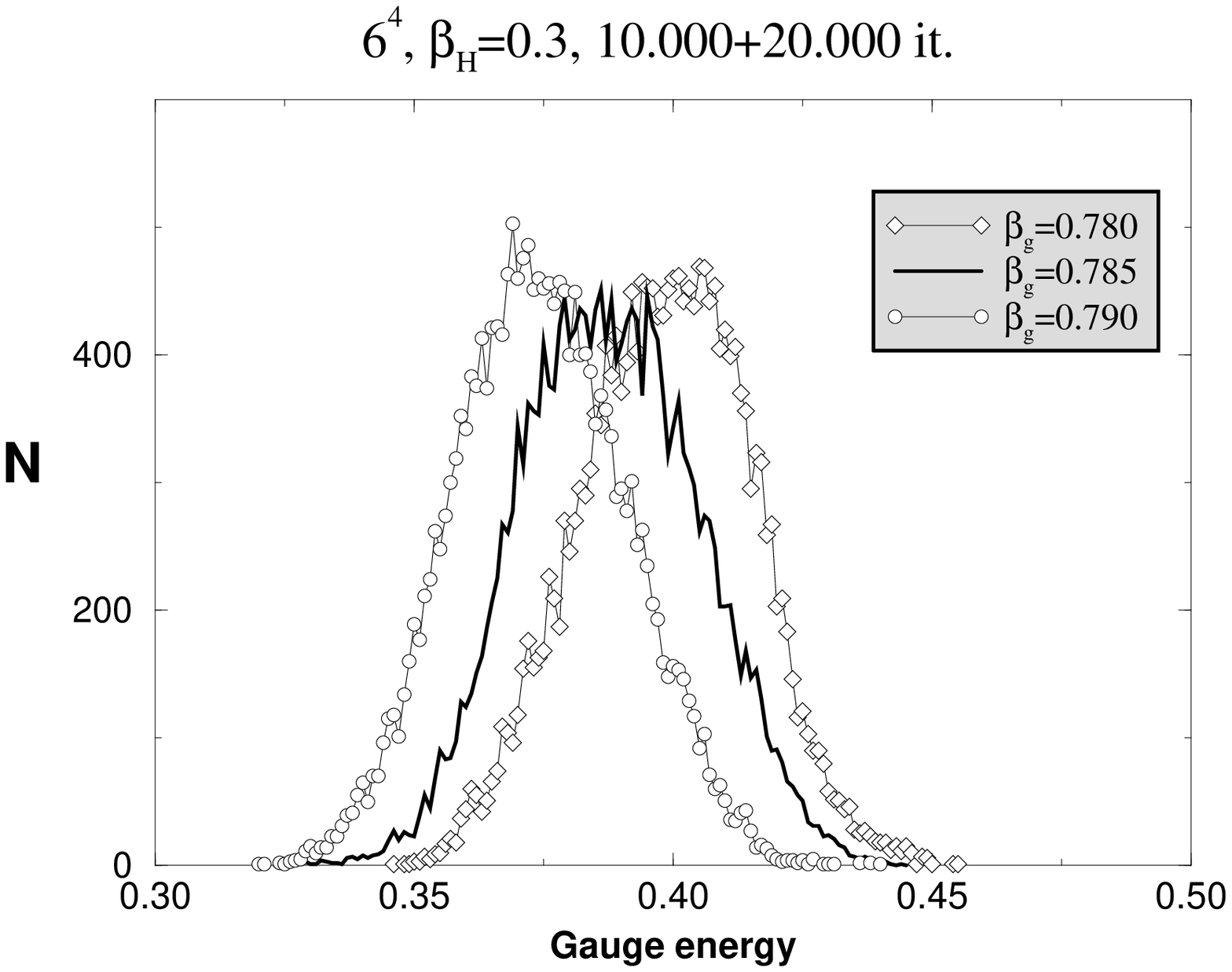}
            {Some histograms. The plot~(a) is evaluated
            in the point $B$.
            Notice the clear two peaks structure revealing a first order
            phase transition. This structure is not present in the 
            plot~(b), evaluated in the point $C$ where no  
            specific heat peak is present.
            }{fig:histo}
}

\newcommand{\figenfunciodebeta}{
\duesfigures{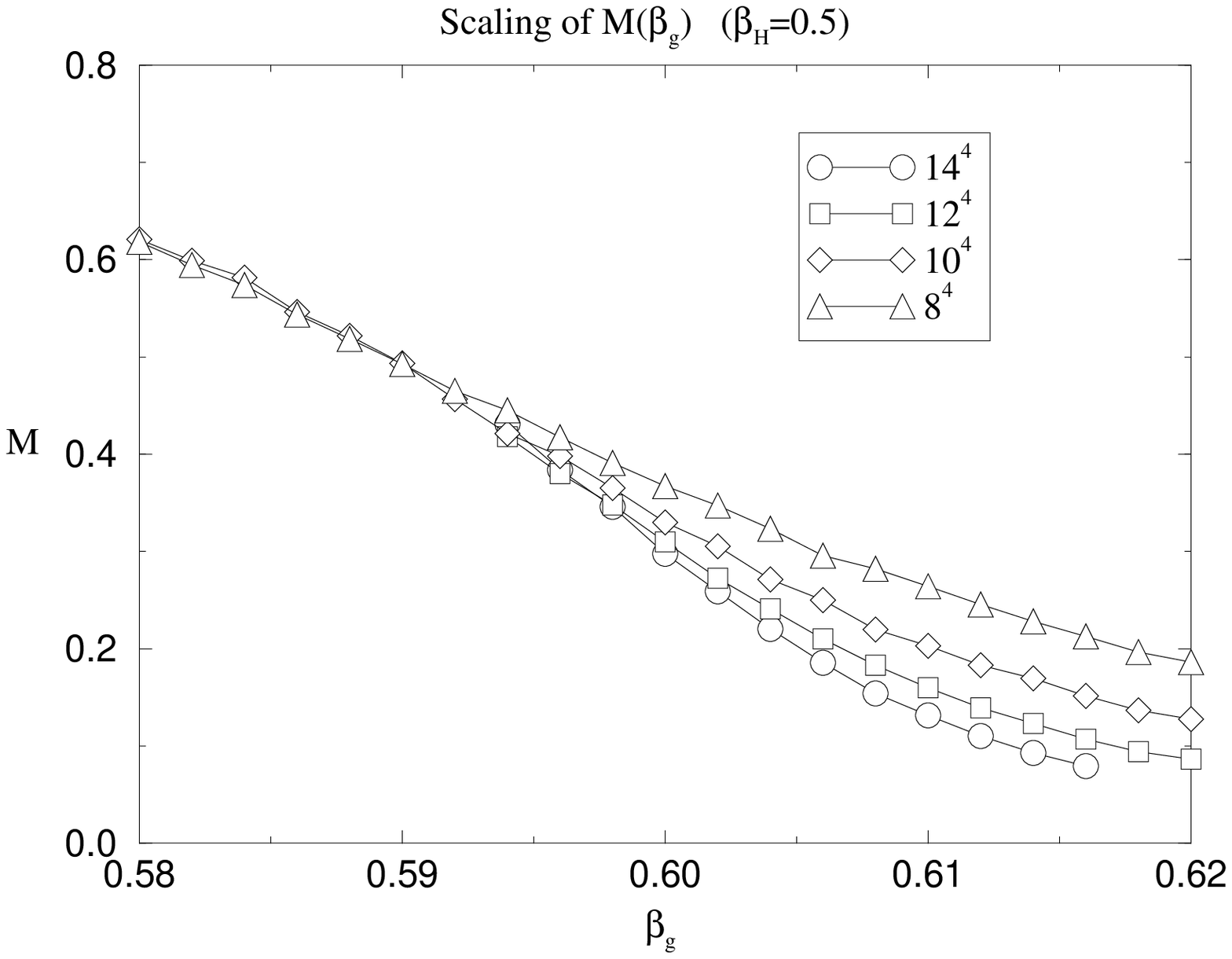}{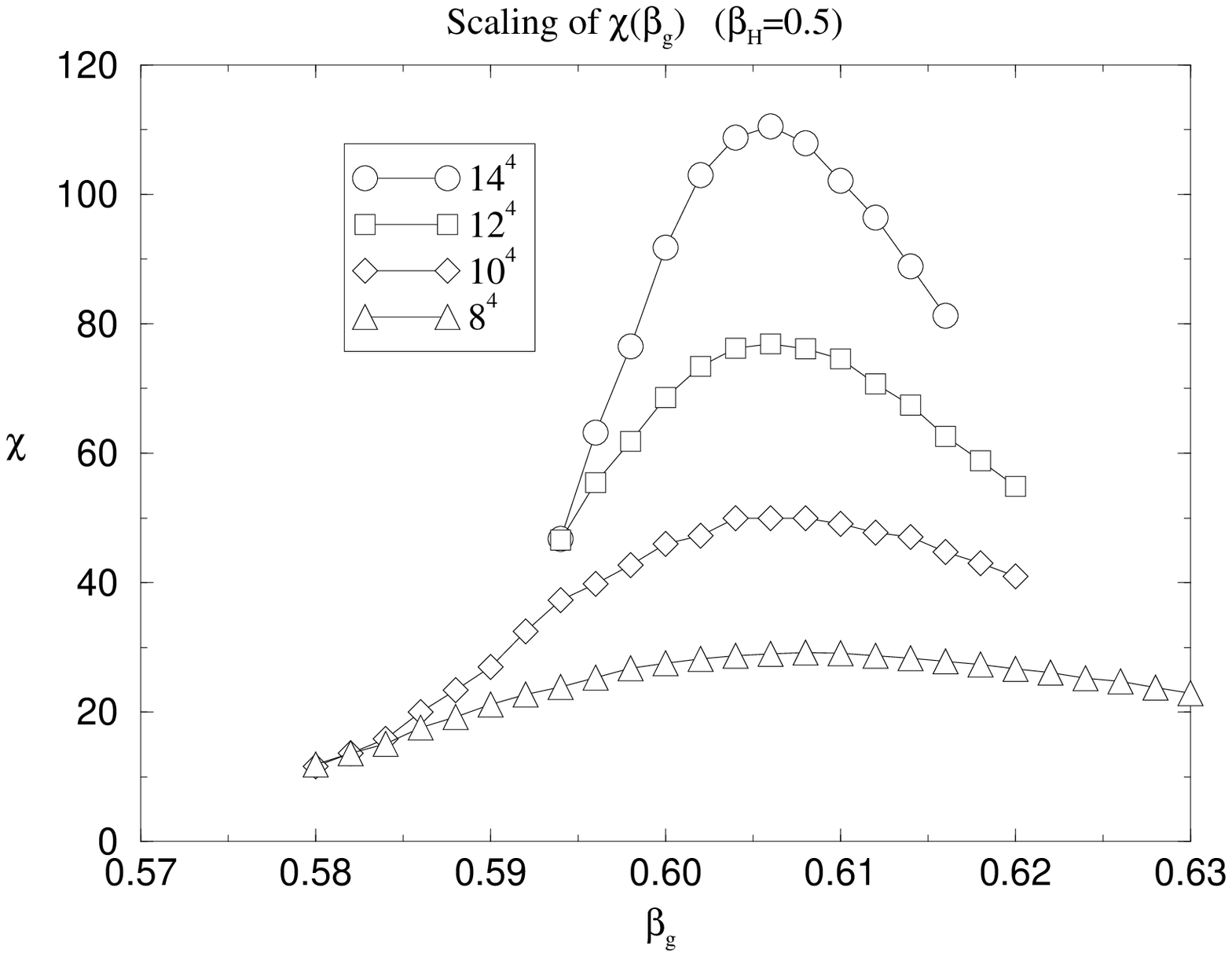}
            {$M$ (in a) and $\chi$ (in b). Both as a function of $\beta_g$,
             keeping $\beta_H=0.5$. The finite size effects of
             $M$ are stronger than in Fig.~\ref{fig:enfuncioderho}a.}
            {fig:enfunciodebeta}
}

\newcommand{\figenfuncioderho}{
\duesfigures{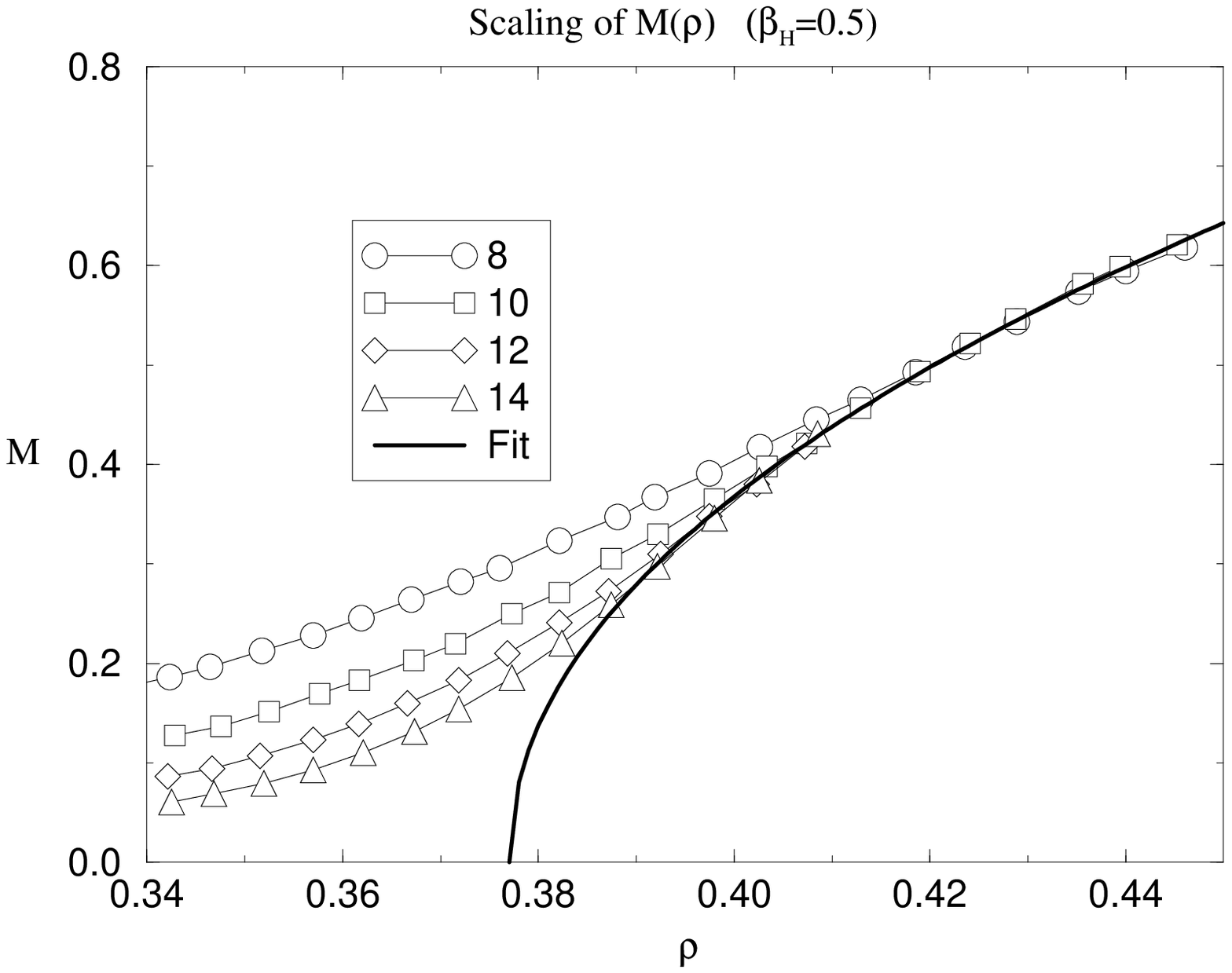}{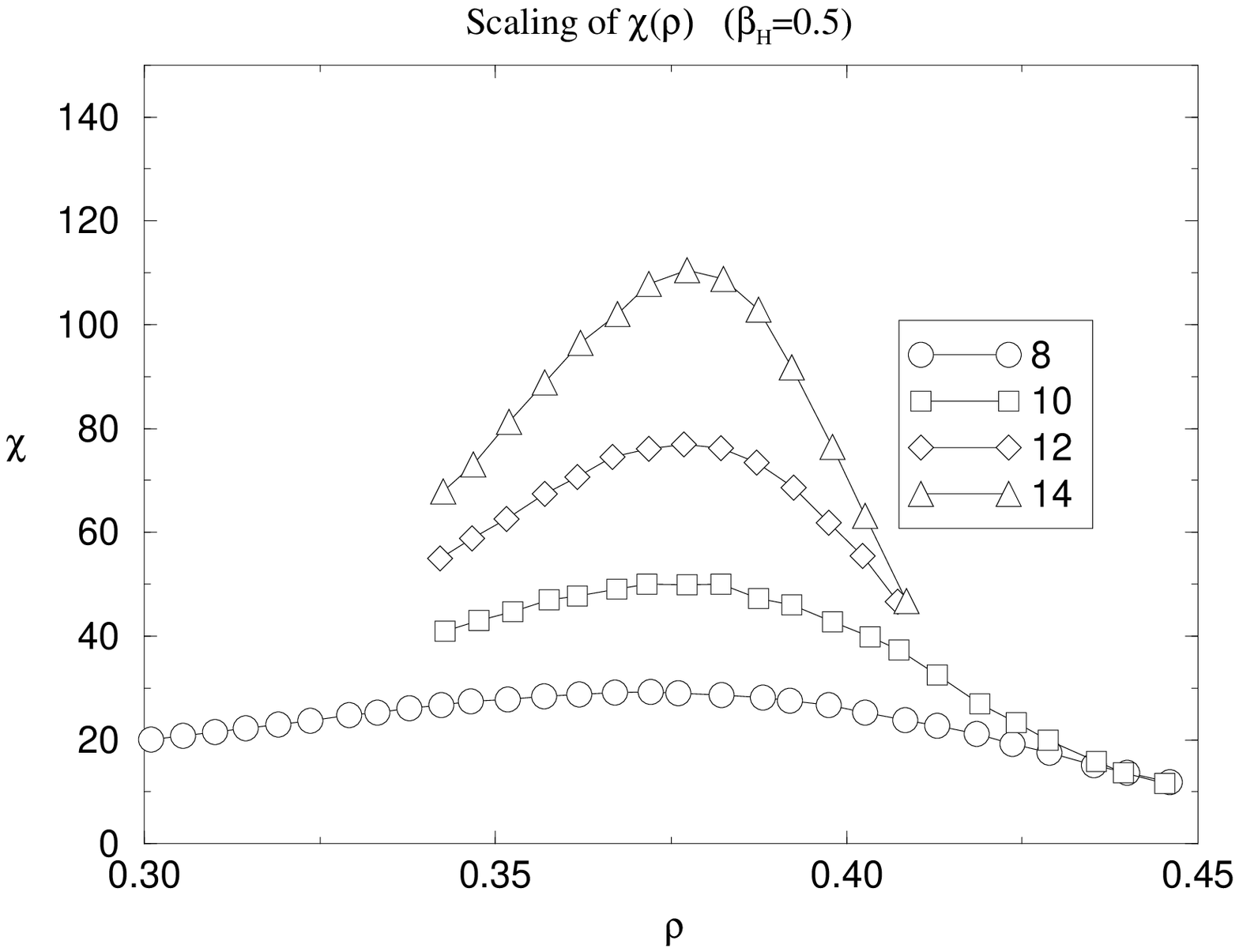}
            {$M$ (in a) and $\chi$ (in b). Both as a function of $\rho$,
             keeping $\beta_H=0.5$. The finite size effects of
             $M$ are lesser than in Fig.~\ref{fig:enfunciodebeta}a.}
            {fig:enfuncioderho}
}

\newcommand{\figfits}{
\duesfigures{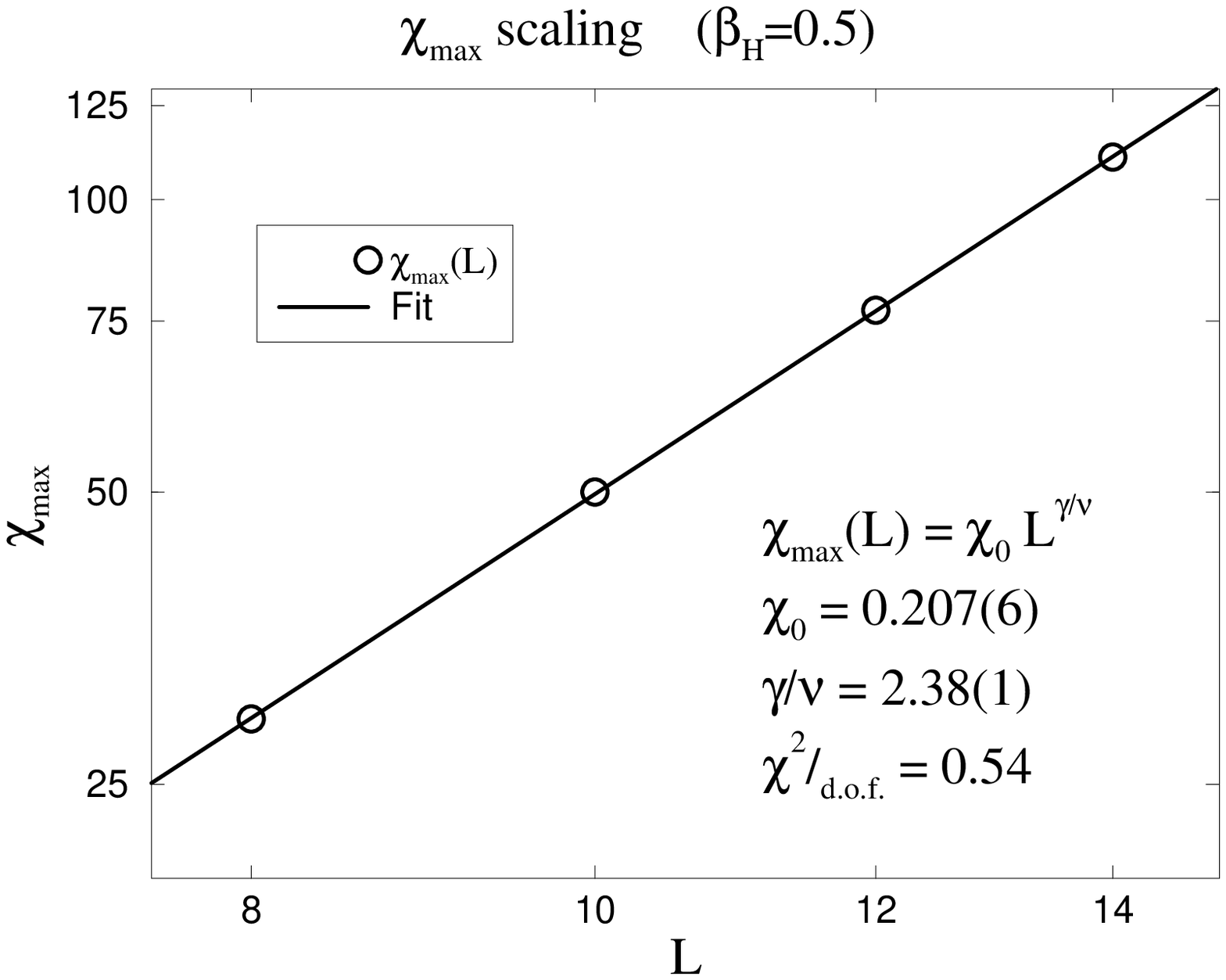}{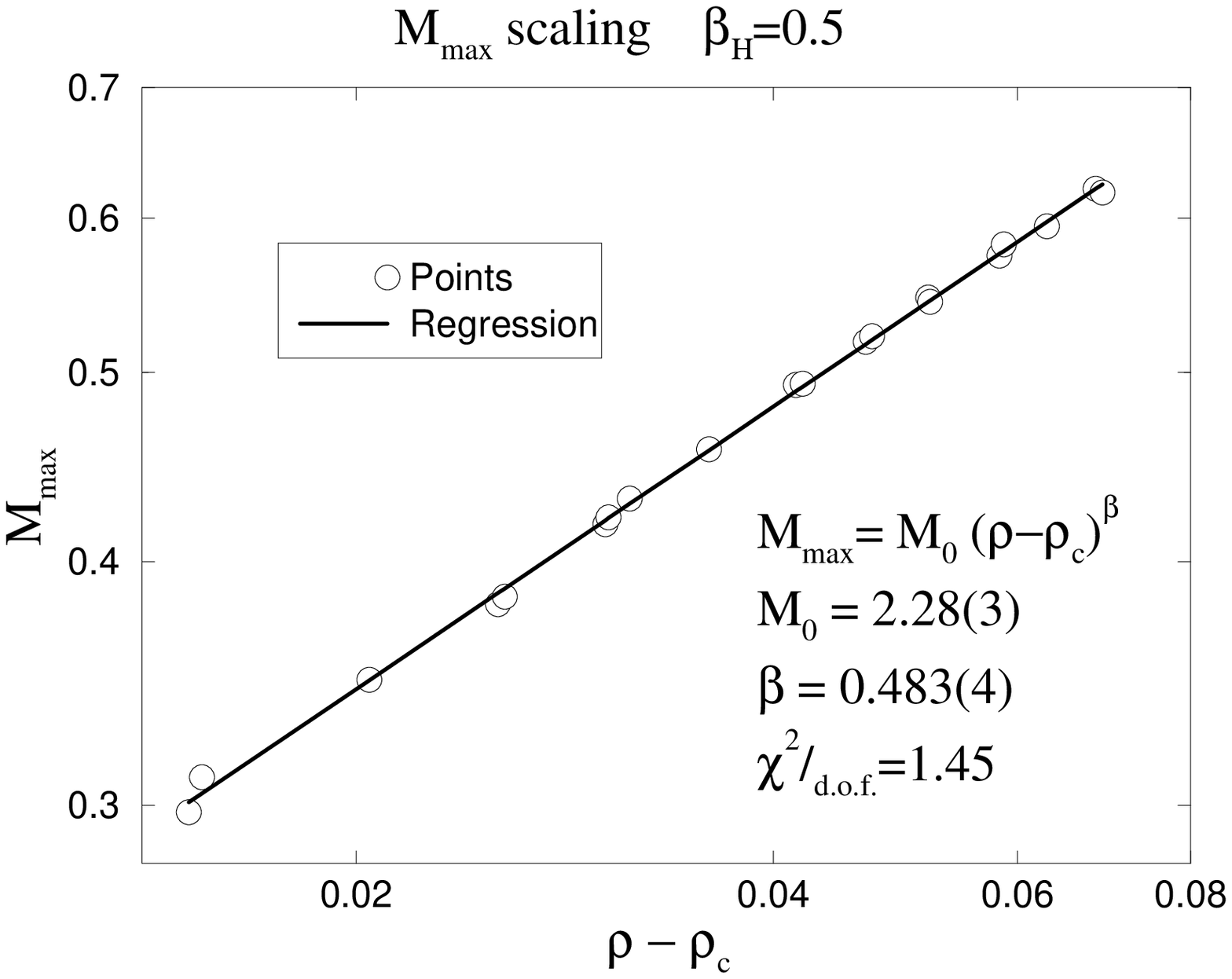}{Log-log plots for 
            $\chi_{max}(L)$ and $M(\rho-\rho_c)$, being $\rho_c=0.377$.
            }{fig:fits}
}

\newcommand{\figfases}{
\duesfigures{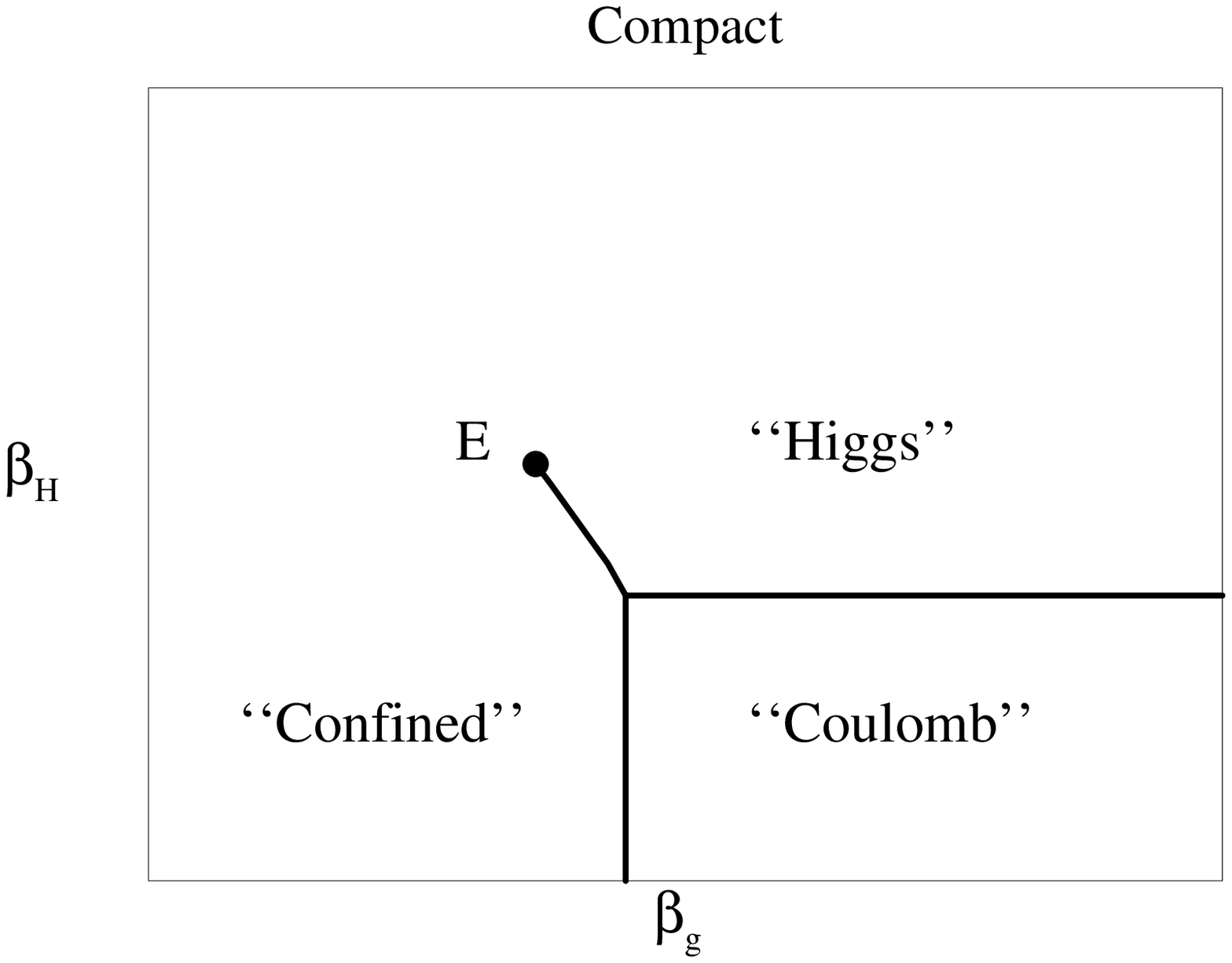}
            {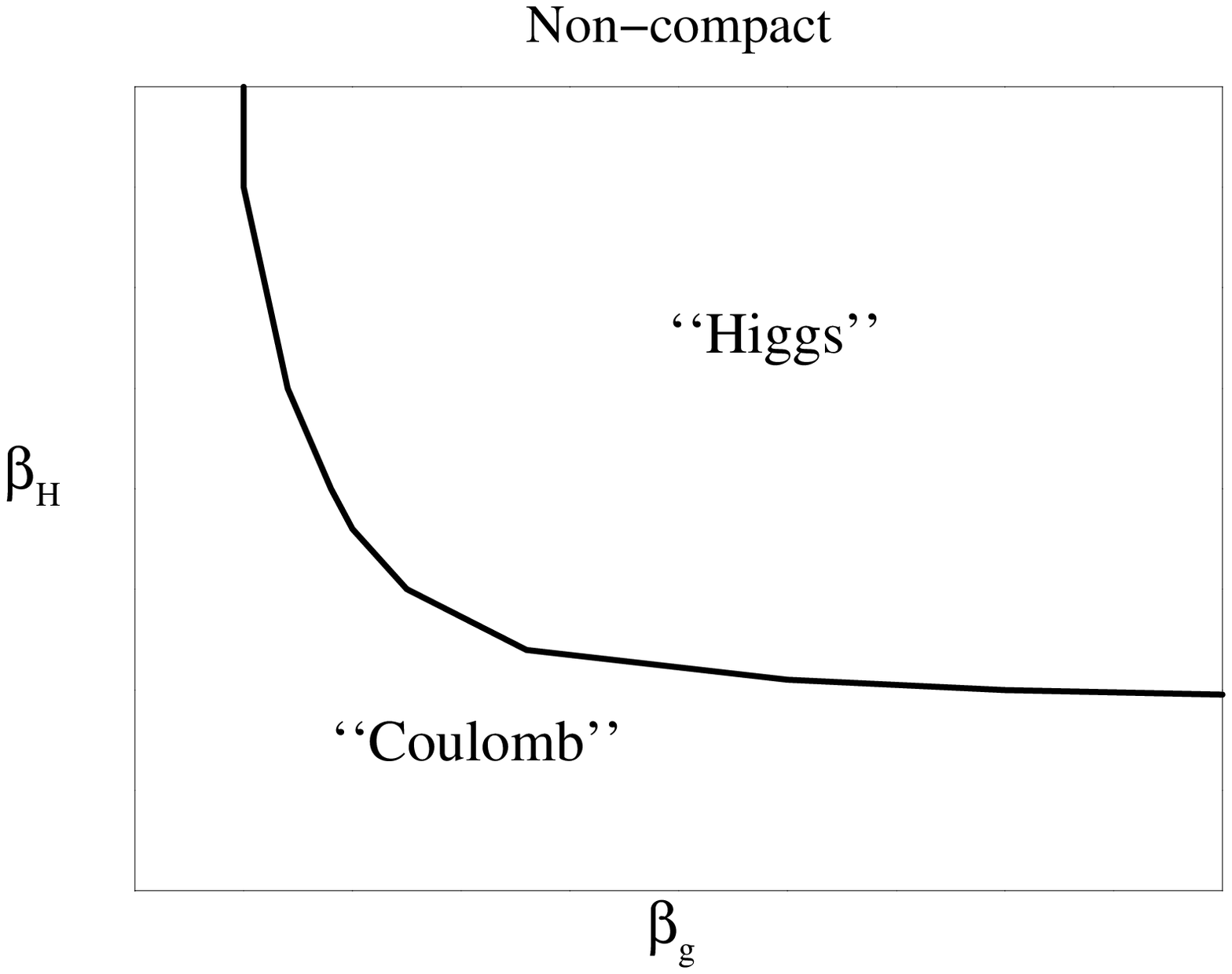}{Qualitative phase diagrams for the
            compact and non-compact scalar QED.
            }{fig:fases}
}

\begin{document}

\draft 
\tightenlines 

\preprint{\vbox{
\hbox{UAB--FT/428}
\hbox{hep--lat/9710042}
}}

\title{Monopole Percolation in the Compact Abelian Higgs Model}

\author{M. Baig and J. Clua}
\address{
Grup de F{\'\i}sica Te{\`o}rica, IFAE\break
Universitat Aut{\`o}noma de Barcelona\break
08193 Bellaterra (Barcelona), Spain.
}

\date{September 20, 1997}
\maketitle

\begin{abstract}
We have studied the monopole-percolation phenomenon in the four dimensional
Abelian theory that contains compact $U(1)$ gauge fields  coupled to 
unitary norm Higgs fields. We have determined the location of the  
percolation transition line in the plane $(\beta_g, \beta_H)$.
This line overlaps 
the confined-Coulomb and the confined-Higgs phase transition lines,
originated by a monopole-condensation mechanism, but continues away from the 
end-point where this phase transition line stops. 
In addition, we have determined the critical exponents of the monopole
percolation transition away from the phase transition
lines. We have performed the finite size scaling in terms of the monopole
density instead of the coupling, because the density seems to be the {\it
natural} parameter when dealing with percolation phenomena. 
\end{abstract}


\section{Introduction}

Much work has been devoted in recent years to the understanding 
of the pure gauge $U(1)$ lattice field theory. Despite the apparent simplicity
of this model, which can be considered a limit of the well known $Z(n)$
gauge models in four dimensions, the order of their  phase transition
is a rather controversial issue. Pioneering simulations suggested that the
transition was actually of second order, {\em i.e.},
a continuum limit for
this lattice theory is possible~\cite{LN80}.
Posterior analysis showed the presence of a two-peak
structure in the histograms  questioning the order of the phase
transition~\cite{JNZ83,ACG91}.
Several new approaches have been proposed:
extended lattice actions~\cite{CF97},
different topologies~\cite{BF94,JL96},
monopole suppressed actions~\cite{KHMM96}, etc.
For a summary of the history of this transition see Ref.~\cite{BSBT95}.

The main point in order to clarify the nature of the phase transition in pure
gauge $U(1)$ lattice field theory is the understanding of
the role of topological
excitations. This analysis was initiated in the early eighties
by several groups: Einhorn and Savit~\cite{ES78} and Banks, Kogut and 
Myerson~\cite{BMK77}.
Their conclusion
was that topological excitations are actually strings of monopole current.
In addition, DeGrand and Toussaint~\cite{DT80} showed that these
monopoles can be actually studied by direct numerical simulation, in
particular the monopole condensation phenomenon over the phase transition. 
On the other hand, it is widely accepted that the observed problems in the 
analysis of the phase transition can be related to the one-dimensional
character of the topological excitations. 

An important observation was made by Kogut, Koci\'c and Hands~\cite{KKH92}.
They showed that in pure gauge non-compact $U(1)$ ({\em i.e.}
the action obtained
keeping only the first order in the Taylor expansion) which is Gaussian,
monopoles percolate and satisfy the hyperscaling relations characteristic
of an actual second-order phase transition. 

Baig, Fort and Kogut~\cite{BFK94} 
showed that in the compact pure gauge theory, just 
over the phase transition point, monopoles not only {\em condensate} but also 
{\em percolate}. They pointed out that the strange behavior of this 
phase transition --its unexpected first order character and strange 
critical exponents-- can be related to the coincidence of these two phenomena.

Furthermore, Baig, Fort, Kogut and Kim~\cite{BFKK95}
showed that in the case of non-compact
QED coupled to scalar Higgs fields, the monopole percolation phenomena 
--previously observed over the gauge line-- actually propagate into the full
$(\beta_g, \beta_H)$ plane,
being $\beta_g$ and $\beta_H$ the gauge and Higgs couplings.

It is surprising that the monopole percolation phenomenon 
is not related to 
the phase transition line that separates the confined and the
Higgs phases, a second order transition that is logarithmically trivial.

In the present paper we have performed a numerical simulation of the compact 
lattice gauge $U(1)$ coupled to Higgs fields of unitary norm.
First of all, we have reproduced some results 
concerning the phase diagram previously obtained by 
Alonso {\it et al.}~\cite{A93}, but 
we have measured
at the same time the behavior of monopoles, condensation and 
percolation. We have determined the location of the 
percolation transition line --the line defined by the maxima of the 
monopole susceptibility--
in the plane $(\beta_g, \beta_H)$. We have observed that this line 
overlaps 
the confined-Coulomb and the confined-Higgs phase transition lines,
which result  by a monopole-condensation mechanism. But it
continues away from the end point where this phase transition line stops. 
In addition, we have determined the critical exponents of the monopole
percolation transition away from the phase transition
lines. We have performed the finite size scaling in terms of the monopole
density instead of the coupling, because the density seems to be the {\it
natural} parameter when dealing with percolation phenomena.

The plan of the paper is the following:
In Sec.~\ref{sec:themodel}
we define our model and we  analyze the appearance of the topological
excitations. 
In Sec.~\ref{sec:percolation}
we extend the standard percolation analysis to the case of
monopole percolation, using densities as critical parameters.
Sec.~\ref{sec:theplane} contains the result of the relation between
monopole percolation and phase transition 
Sec.~\ref{sec:finitesizescaling} is devoted to the finite size analysis
of the percolation phenomena.
Finally, Sec.\ref{sec:conclusions}
contains the conclusions of this work.\footnote{
While our work was being completed, it appeared a work by
Franzki, Kogut and Lombardo~\cite{FKL97} where they analyze
a model with both, scalars and fermions. In the conclusions 
we compare  the results coming from both simulations and, in particular,
we discuss the different ways of extracting the critical exponents.}

\section{The Model}
\label{sec:themodel}

\subsection{The action}
In this paper, we resume  the work initiated in
Ref.~\cite{BFK94}, and focus on the compact abelian Higgs Model 
with a fixed length scalar field whose action is

\begin{equation}
S=-\beta_g\sum_{n\mu\nu}\cos \Theta_{\mu\nu}(n)
-\beta_H\sum_{x,\mu} \left( \phi_x^*U_{x,\mu}\phi_{x+\mu} +c.c. \right),
\label{eq:compactaction}
\end{equation}
where  $\Theta_{\mu\nu}(n)$
is the circulation of the compact
gauge field around a plaquette, $\beta_g=1/e^2$
is the gauge coupling
and $\phi_x=\exp{i\alpha(x)}$ is a phase factor. We choose fixed
length Higgs fields because it has been shown that they live in the same
universality class than conventional variable length Higgs fields but have the
advantage that they do not require fine-tuning~\cite{CP86}.
In addition, we want to  keep
as close as possible to
the analysis of the non-compact Abelian Higgs
Model of Refs.~\cite{BFKK95,BDKM90} in order to compare to
non-compact scalar
electrodynamics, whose action is
\begin{equation}
S_{nc}=\frac{1}{2} \beta_g \sum_{n\mu\nu} \Theta_{\mu\nu}(n)^2
-\beta_H\sum_{x,\mu} \left( \phi_x^*U_{x,\mu}\phi_{x+\mu} +c.c. \right).
\end{equation}
The phase diagrams for the two models are well known. They are qualitatively
represented in Fig.~\ref{fig:fases}.
Non-compact scalar QED has two disconnected phases: a Coulomb
phase and a Higgs phase. It is important to remark that there is not phase
transition over the pure gauge axis.
In contrast, in the compact case a
transition separates the Coulomb and the confined phase while the
transition separating the confined and Higgs phases has an end point 
(point $E$ in the Fig.~\ref{fig:fases}a).

\figfases

\subsection{Monopoles}
Following~\cite{DT80} we can separate the plaquette angle 
$\Theta_{\mu\nu}$ into
two pieces: physical fluctuations which lie in the range $-\pi$ to 
$\pi$ and 
Dirac strings which carry $2\pi$  units of flux. Introducing an 
electric 
charge $e$ we define an integer-valued Dirac string by
\begin{equation}
e\Theta_{\mu\nu}=  e\bar\Theta_{\mu\nu}(\tilde n) + 2\pi S_{\mu\nu},
\end{equation}
where the integer   $S_{\mu\nu}$   determines the strength of the 
string threading the plaquette and  $e\Theta_{\mu\nu}$ represents 
physical fluctuations. 
The integer-valued monopole current, $m_{\mu}(n)$, defined 
on links of the dual lattice, is then
\begin{equation}
m_\mu(\tilde n)= {1\over 2} \epsilon_{\mu\nu\kappa\lambda}
\Delta_\nu^+ S_{\kappa\lambda}(n+\hat\mu),
\label{eq:monopolecurrent}
\end{equation}
where $\Delta_\nu^+$ is the forward lattice difference operator, and
$m_\mu$ is the
oriented sum of the $S_{\mu\nu}$ around the faces of an elementary 
cube. This gauge-invariant definition implies the conservation law
$\Delta_\mu^- m_\mu(\tilde n)=0$,
which means that monopole world lines form closed loops.

The appearance of topological excitations in the abelian Higgs model was
studied in Ref.~\cite{ES78,BMK77}. Furthermore, in Ref~\cite{RKR82} a Monte
Carlo simulation was performed to look for the density of monopoles. They
observed that the 
confined (Higgs and Colulomb) phase was characterized by a 
large (vanishing) 
density of monopoles vortices and electric current densities.

\section{Percolation}
\label{sec:percolation}

\subsection{Concepts of percolation}

The role of the monopole percolation in this model is analyzed using the techniques
of standard percolation~\cite{Sta79}. In the simplest models 
the sites are occupied with a probability $p$.
One can define a cluster as a group of occupied sites connected by
nearest-neighbor distances.
When a cluster becomes infinite in extent, it is called a
percolation cluster. Obviously, for $p=0$ all the sites are empty and there 
is no percolation cluster.  For $p=1$ all the sites are occupied and
{\em one} percolation cluster exists.
There exists a critical concentration $p_c$
such that for $p<p_c$ ($p>p_c$) no (one) percolation cluster exists.
This is the typical behavior of a phase
transition. We define $P_\infty$ as the probability for an occupied site
to belong to the infinite cluster.

If we define $g_n$ as the relative number of clusters of
size $n$, the probability of any site to belong to a cluster of size $n$ is
$n g_n$.
According to this definitions, the mean cluster size is

\begin{equation}
               S = \frac{\sum_n n^2  g_n}{\sum_n n  g_n},
\end{equation}
where the infinite cluster is excluded from the sum.
This quantity diverges at the critical concentration.

Thus, $P_\infty$ is the order parameter of the transition and $S$ is its
associated susceptibility. Their behaviors near the critical point are:

\begin{equation}
         P_\infty  \sim  \left( p - p_c \right)^\beta      \\
\hspace{1cm} \mathrm{for} \hspace{12pt} p > p_c,
\end{equation}
and
\begin{equation}
         S  \sim  \left| p - p_c \right|^{-\gamma}.
\end{equation}

\subsection{Percolation of monopoles in QED}
We have defined 
the monopole current $m_\mu(x)$ in eq.~(\ref{eq:monopolecurrent}).
One can define a connected cluster of monopoles as a set of sites joined by
monopole line elements~\cite{HW89}. Notice that this definition ignores
the fact that monopoles are actually vectors.
The density of occupation is
\begin{equation}
  \rho = \frac{n_{tot}}{L^4} = \sum_{n=4}^{n_{max}} g_n n,
\label{eq:rho}
\end{equation}
where $n_{tot}$ is the total number of connected sites and $n_{max}$
is the size of
the largest cluster. The number~4 in the sum comes from the conservation law.

The order parameter ($P_\infty$) is
\begin{equation}
 M = \frac{n_{max}}{n_{tot}}.
\label{eq:M}
\end{equation}

Its corresponding susceptibility (the mean cluster size $S$) is
\begin{equation}
  \chi = \frac{1}{n_{tot} } 
  \left( \sum_{n=4}^{n_{max}} g_n n^2 -n^2_{max} \right).
  \label{eq:chi}
\end{equation}

Near the critical point, $M$ should have a non analytical behavior with
a ``magnetic''
exponent $\beta$. We have followed lines of constant $\beta_H$, so
\begin{equation}
M \sim \left( \beta_g^{c} - \beta_g\right)^\beta 
\hspace{1cm} \mathrm{for} \hspace{12pt}
\beta_g < \beta_g^{c}.
\end{equation}
where the upperscript $c$ means ``critical''.
The susceptibility also diverges:
\begin{equation}
\chi  \sim \left( \beta_g^{c} - \beta_g\right)^{-\gamma}.
\end{equation}

One should remark that the {\it natural} parameter when studying percolation 
is the probability of occupation (in random percolation it determines all the
observables).
In a lattice gauge model, the fundamental variables
are the couplings, which determine all the other observables. In particular,
they determine the density of monopoles $\rho$,
{\em i.e.}, the probability of occupation. In this sense, it is also possible
to parameterize percolation as a function of the densities.
In this case the critical behavior will be
\begin{equation}
M \sim \left( \rho - \rho_c \right)^\beta
\hspace{1cm} \mathrm{for} \hspace{12pt}
\beta_g < \beta_g^{c}, \label{eq:beta-density}
\end{equation}
and
\begin{equation}
 \chi  \sim \left( \rho - \rho_c \right)^{-\gamma}.
\end{equation}
Clearly, the two parameterizations might have the same critical exponents,
but the first one seems to exhibit stronger finite size effects.

\section{Numerical results}

\subsection{Phase transition {\em vs.} monopole percolation}
\label{sec:theplane}

We have performed numerical simulations with the
action~(\ref{eq:compactaction}) on hypercubical lattices 
with standard periodic boundary conditions. In these simulations we have
measured the internal gauge and Higgs energies
\begin{equation}
E_g = \sum_{n\mu\nu}\cos \Theta_{\mu\nu}(n),
\end{equation}
\begin{equation}
E_H = \sum_{x,\mu} \left( \phi_x^*U_{x,\mu}\phi_{x+\mu} +c.c. \right).
\end{equation}
as well as all magnitudes related to monopole percolation, {\em i.e.} 
$\rho$, $M$ and $\chi$, which are defined in 
eqs.~(\ref{eq:rho},~\ref{eq:M},~\ref{eq:chi}) respectively.

In order to locate the continuation into the full $(\beta_g, \beta_H)$
plane of
the monopole percolation transition previously determined over the pure
gauge axis in Ref~\cite{BFK94}, we have performed repeated measurements
over thermal cycles 
changing the $\beta_g$ coupling, until the entire region inside 
the limits $0<\beta_g<1.5$ and $0<\beta_H<1.5$  has been covered.
In particular, the confined-Coulomb
and confined-Higgs phase transition lines. Lattice size for this exploratory
runs has been of $6^4$ and the $\beta_g$ step has been
$\Delta\beta_g=0.005$. 

We have not used any
reweighting technique when measuring the monopole-related magnitudes. 
In all the plots, lines are only to guide the eyes and the errors are not
shown because they are smaller than the symbols used.

The number
of iterations at each $\beta_g$ value has been $30.000$, discarding the first
$10.000$. In addition, several thermal cycles 
have been performed at different values of $\beta_H$
in order to determine the location of the Coulomb-Higgs phase transition line.

Fig.~\ref{fig:plane} collects the results of these runs.
The small black points 
represent the maximum of the monopole susceptibility. 
The solid line is the phase-transition line.
Note that the monopole percolation transition line is on top of both the
confined-Coulomb line and  the confined-Higgs line up to the end point.
Beyond this point monopole percolation transition line continues, 
now unrelated to any energy phase transition, approaching the vertical axis.
This
behavior is in clear contrast with the non-compact case~\cite{BFKK95} where the
monopole-transition is completely independent of  the Coulomb-Higgs phase 
transition.

\figplane 

In Fig.~\ref{fig:histo} we compare two sets of histograms corresponding to
cycles:
(a) for $\beta_H=0.25$, over the confined-Higgs phase transition
line and (b) for $\beta_H=0.3$, {\em i.e.} just beyond the end-point $E$.
In the two
figures the solid line corresponds to the value of $\beta_g$ for which the
monopole susceptibility has a maximum. The appearance of a two-peak structure
over the phase transition confirms that the lattice size, the statistics
used and the coupling step in the thermal cycle are enough to establish the
appearance of the phase transition.

\fighisto 

An interesting question to establish is the concordance of all the
phenomena (phase transition, monopole condensation and monopole
percolation) at the same coupling.
In this sense it was proposed in Ref.~\cite{BFK94} to use $M$ 
as an order parameter to determine the location of the phase transition.
The existence of an end-point for the confined-Higgs phase transition
line gives us a chance to check the behavior of these three
parameters over and right beyond the end-point.
In Fig.~\ref{fig:dades} we have collected the results for
thermal cycles at different values of $\beta_H$. Fig.~\ref{fig:dades}a
corresponds to the pure gauge compact action (point $A$ in the
phase diagram), the case of Ref.~\cite{BFK94}.
Fig.~\ref{fig:dades}b corresponds to $\beta_H=0.25$, {\em i.e.}, over the
confined-Higgs phase transition (point $B$).
The behavior of all three parameters is similar to that observed in
the pure gauge case, although the discontinuity is more abrupt.
Fig.~\ref{fig:dades}c corresponds to $\beta_H=0.30$ (point $C$),
just after the end
point, and Fig.~\ref{fig:dades}d to $\beta_H=0.5$ (point $D$),
a line that crosses the
maximum of the monopole susceptibility but  is far away from the end-point.
In these last cases the energy clearly shows no discontinuity
and the monopole density is completely smooth.
Nevertheless, the $M$ parameter shows a fast decrease
corresponding to the percolation phenomena 
investigated in the next subsection.

\figdades 

Fig.~\ref{fig:peaks} shows another hint of the change of nature of
the percolation at the end-point. We show the maximum of $\chi$ along the
percolation line. A clear change can be seen  at the end point.

\figpeaks

\subsection{Finite size scaling of monopole percolation transition}
\label{sec:finitesizescaling}

We want to determine the critical exponents of the monopole percolation
away from the phase transition lines, at the point
$D=(0.5, 0.606)$ in the plane $(\beta_g, \beta_H)$.
To do this, we keep 
$\beta_H=0.5$ fixed, and we vary $\beta_g$ within the range $[0.594, 0.620]$
for lattice sizes ranging from $8^4$ to $12^4$.
We have recorded some percolation
parameters for each simulation: the density of monopoles $\rho$,
$M$ and $\chi$.
The statistics for each point is of
$10.000$ iterations once the first $5.000$ have been discarded.

In Fig.~\ref{fig:enfunciodebeta},
we collect the results for $M$ and $\chi$ as a function of $\beta_g$.
\figenfunciodebeta
In Fig.~\ref{fig:enfuncioderho}, we show the same results for $M$ and $\chi$
as a function of $\rho$.
\figenfuncioderho

According to  finite size scaling arguments~\cite{Bar83},
the peak of the susceptibility
should grow with the lattice size $L$ as
\begin{equation}
 \chi_{max} \sim L^{\gamma/\nu}, \label{eq:chi-scaling}
\end{equation}
and the value of $M$ at the critical point must vanish as
\begin{equation}
 M_{crit} \sim L^{-\beta/\nu}. \label{eq:M-scaling}
\end{equation}

\figfits

We show the value of $\chi_{max}$ as a function of $L$ in a log-log plot in
the Fig.~\ref{fig:fits}a. Notice that the
scaling law is satisfied. A fit to eq.~(\ref{eq:chi-scaling}) gives
\begin{equation}
        \frac{\gamma}{\nu} = 2.38(1).
\end{equation}
The straight line resulting from the fit is also shown.

We measure $\beta$ fitting the values that are not distorted by
finite-size effects to the equation~(\ref{eq:beta-density}). 
The result is
\begin{equation}
        \beta = 0.483(4).
        \label{eq:beta}
\end{equation}
In Fig.~\ref{fig:fits}b we show the selected points and
the  fit, with a value for the critical density about of $\rho_c=0.377$.

The fit of~(\ref{eq:M-scaling}) is not very good.
It gives $\beta/\nu=0.7(1)$.
Putting our measurements of $\gamma/\nu$
into the hyperscaling relation
\begin{equation}
  \frac{\beta}{\nu} = \frac{1}{2} \left( d - \frac{\gamma}{\nu}\right),
\end{equation}
we obtain 
\begin{equation}
  \frac{\beta}{\nu}=0.81(1),
  \label{eq:beta/nu}
\end{equation}
which is
more accurate than our measurement. So, we take this value and regard
the measurement as a test of hyperscaling.

Combining eqs.~\ref{eq:beta}~and~\ref{eq:beta/nu}, the value of $\nu$ is
\begin{equation}
        \nu = 0.60(1).
\end{equation}

The remaining critical exponents can be derived from those above using
hyperscaling relations.

In the below table we compare our results with other related works:

\vspace{0.30cm}
{\centering \begin{tabular}{|c|c|c|c|}
\hline 
&\( p_c \)&\( \beta  \)&\( \nu  \)\\
\hline 
\hline 
Our results&0.377&0.483(4)&0.60(1)\\
\hline 
Non compact quenched QED~\cite{KKH92}&-&0.58(2)&0.66(3)\\
\hline 
Pure four dimensional site percolation~\cite{BFMSPR96}&0.161&0.715&0.683\\
\hline 
FKL97~\cite{FKL97}&-&0.50(4)&0.61(4)\\
\hline
\end{tabular}\par}
\vspace{0.30cm}

The vector nature of the monopole current suggests that our model do not
lie in the same universality class that the four-dimensional site percolation.

We also see that our values are closed to the non-compact QED~\cite{KKH92},
but more data is required to decide if the two models belong to the same
universality class.

Finally, we would like to stress the remarkable agreement with~\cite{FKL97}.
Nevertheless,
they perform their analysis using a different value of the Higgs
coupling, $\beta_H \sim 0.9$.
As it pointed out in~\cite{FKL97},
a more careful analysis is necessary to decide if the 
critical exponents are the same along the percolation line 
beyond the end point.

\section{Conclusions}
\label{sec:conclusions}

We have performed a numerical simulation of the compact 
lattice gauge $U(1)$ coupled to Higgs fields of unitary norm.
Some results 
concerning the phase diagram previously obtained by 
Alonso {\it et al.}~\cite{A93}, 
have been confirmed 
but measuring at the same time the behavior of monopoles, condensation and 
percolation. 
The location of the percolation transition line
--the line defined by the maxima of the monopole susceptibility--
in the plane $(\beta_g, \beta_H)$ has been determined.
This line
overlaps 
the confined-Coulomb and the confined-Higgs phase transition lines,
which are originated by a monopole-condensation mechanism, but it
 continues away from the end-point.
This is in contrast with the behavior of the monopole percolation transition
in the non-compact abelian Higgs model where it is unrelated to any phase
transition.
In addition, the critical exponents of the monopole
percolation transition in the region far away from the phase transition
lines have been determined. The finite size scaling has been performed in 
terms of the monopole density instead of the coupling, because the density 
seems to be the {\it natural} parameter when dealing with percolation 
phenomena.

While this paper was written up we have received
Ref.~\cite{FKL97} where a model with both, scalar and fermion matter fields
has been considered. Their results for the scalar sector are in perfect
agreement with our simulations. It is interesting to compare the finite size
scaling analysis for the critical exponents
in the pure-percolation region obtained
using couplings and density parameterizations.
The values obtained in both cases are in perfect agreement.
However our density parameterization results have been obtained with
lattice sizes smaller than those in Ref.~\cite{FKL97}.
This suggests that the
finite volume effects are smaller if our parameterization is used.

\acknowledgments
Part of the numerical computations have been performed in CESCA (Centre de
Supercomputaci{\'o} de Catalunya). This work has been partially supported by
research project CICYT AEN95/0882. We thank Emili Bagan.

\end{document}